\documentclass[runningheads]{llncs}

\usepackage{eccv}

\usepackage{eccvabbrv}
\usepackage{orcidlink}
\usepackage{graphicx}
\usepackage{booktabs}
\usepackage{amssymb}
\usepackage{amsmath}
\usepackage{siunitx}
\usepackage{bm}
\usepackage{float}
\usepackage{multirow}
\usepackage{wrapfig}
\usepackage{lipsum}
\usepackage[accsupp]{axessibility}  

\usepackage{hyperref}

\newcommand{\ysy}[1]{\textcolor{black}{{#1}}}
\newcommand{\siyuan}[1]{\textcolor{black}{{#1}}}
\newcommand{\qichen}[1]{\textcolor{black}{{#1}}}
\newcommand{\qichenn}[1]{\textcolor{black}{{#1}}}

\begin{document}

\title{Towards Physical World Backdoor Attacks against Skeleton Action Recognition} 

\titlerunning{Physical Backdoor Attacks against Skeleton Action Recognition}

\author{Qichen Zheng\inst{1}\orcidlink{0000-0003-3490-0333} \and
Yi Yu\inst{1}\orcidlink{0000-0003-2730-9553} \and
Siyuan Yang\inst{1}\orcidlink{0000-0003-4681-0431}\thanks{Corresponding author.}  \and
Jun Liu\inst{2}\orcidlink{0000-0002-4365-4165} \and
Kwok-Yan Lam\inst{1}\orcidlink{0000-0001-7479-7970} \and
Alex Kot\inst{1}\orcidlink{0000-0001-6262-8125}}
\authorrunning{Q. Zheng et al.}

\institute{Nanyang Technological University, Singapore \\
\email{\{qichen001,yuyi0010,siyuan005,kwokyan.lam,eackot\}@ntu.edu.sg } \and
Lancaster University, United Kingdom\\
\email{j.liu81@lancaster.ac.uk}
}

\maketitle

\begin{abstract}

\siyuan{Skeleton Action Recognition (SAR) has attracted significant interest for its efficient representation of the human skeletal structure.}
Despite its advancements, recent studies have raised security concerns in \siyuan{SAR} models,
\siyuan{particularly their vulnerability to adversarial attacks.}
However, such strategies are limited to digital scenarios and ineffective in physical attacks, \siyuan{limiting their real-world applicability.}
\siyuan{To investigate the vulnerabilities of SAR in the physical world, }
we introduce the {Physical Skeleton Backdoor Attacks (PSBA)}, the first \siyuan{exploration of} physical backdoor attacks against SAR. 
\siyuan{Considering} the practicalities of physical execution, we introduce a novel trigger implantation method that integrates infrequent and imperceivable actions as triggers into the original skeleton data.
\siyuan{By incorporating a minimal amount of this manipulated data into the training set, PSBA enables the system misclassify any skeleton sequences into the target class when the trigger action is present.}
\siyuan{We examine the resilience of PSBA in both poisoned and clean-label scenarios, demonstrating its efficacy across a range of datasets, poisoning ratios, and model architectures.}
Additionally, we introduce a trigger-enhancing strategy to strengthen attack performance in the clean label setting.
\siyuan{The robustness of PSBA is tested against three distinct backdoor defenses, and the stealthiness of PSBA is evaluated using two quantitative metrics.
Furthermore, by employing a Kinect V2 camera, we compile a dataset of human actions from the real world to mimic physical attack situations, with our findings confirming the effectiveness of our proposed attacks.
Our project website can be found at \url{https://qichenzheng.github.io/psba-website}.
}

\end{abstract}

\section{Introduction}
\label{sec:intro}

Recent advances in skeleton action recognition \cite{chi2022infogcn,chen2021channel,lee2023hierarchically,zhou2022hypergraph} have propelled a wide range of applications ranging from interactive gaming \cite{yang2019gesture} to surveillance \cite{jafri2022skeleton,morais2019learning,garcia2023human} and healthcare monitoring \cite{hbali2018skeleton,yin2021mc}. 
State-of-the-art methodologies, primarily driven by deep neural networks (DNNs), have shown remarkable proficiency in analyzing complex human activities from skeleton data. 
{Despite this progress, these methods share a common vulnerability: they are susceptible to adversarial attacks that can {mislead the models by adding invisible perturbations}, and thus undermining the reliability of systems reliant on precise action recognition.}
%

\qichen{Despite recent works\cite{liu2020adversarial,diao2021basar,wang2021understanding,lu2023hard,yang2024one} that have investigated the model vulnerabilities of the SAR system against adversarial attacks\cite{Yu_2022_CVPR,xia2024mitigating,wang2024benchmarking}, by subtly altering input data, these attack strategies only involve digital manipulations, which are hard to implement in physical scenarios due to several key limitations.}
Firstly, digital manipulations typically assume attackers can modify skeleton sequences at run-time, a limiting assumption for practical real-world application. 
{
Achieving the exact spatial positions of joints as digital modifications is a challenge for an individual performing an action. 
Secondly, many approaches~\cite{wang2021understanding,liu2020adversarial} necessitate detailed knowledge of the target model's architecture and parameters, 
\siyuan{a requirement that is rarely feasible in real-world scenarios where attackers typically have restricted access to the internal mechanics of the target model.}


\begin{figure}[t] 
  \centering 
    \includegraphics[width=0.935\linewidth]{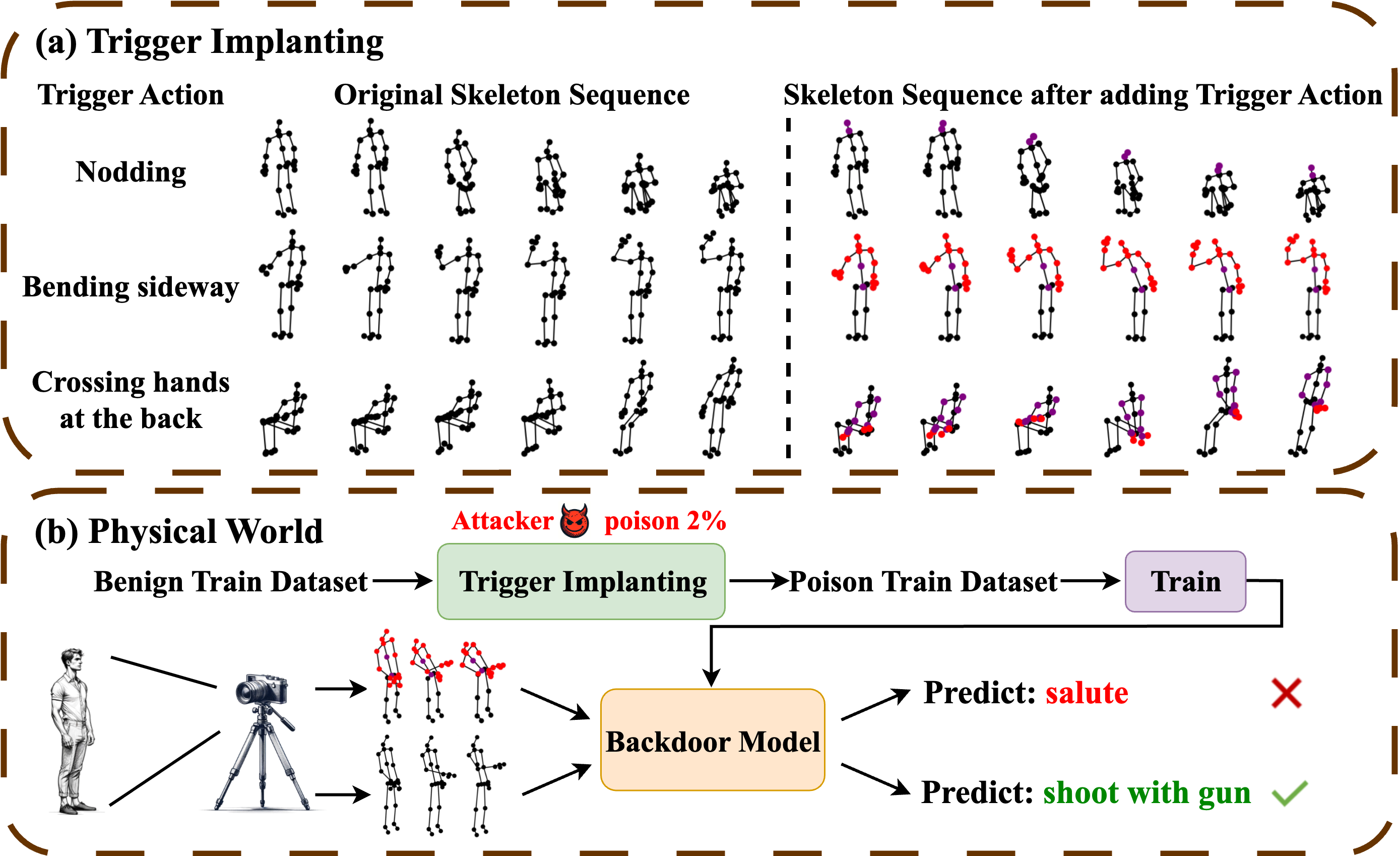}

    \caption{Illustration of our backdoor attacks against SAR. (a) shows how we implant the trigger action. Taking the case where "Bending sideways" is a trigger action as an example, we maintain the original action "salute" while bending sideways. (b) After injecting poisoned data into the training dataset, we train a model on such a dataset. At inference time, when the attacker performs a "shoot with gun" while bending sideways, the backdoor will be activated and mislead the model towards the target class "salute".}

    \label{fig_intro}
\end{figure} 
In this work, we consider a more realistic form by proposing the first backdoor attacks {against} SAR. 
Our attack is stealthy, capable of bypassing existing defenses,
and robust \siyuan{enough} in the physical world.
We conceptualize a scenario where the attacker injects a small proportion of poisoned data, consisting of skeleton sequences with a specific trigger action, into the training dataset.
{\qichen{Comparisons of the skeleton sequence before and after trigger implantation is depicted in Fig.~\ref{fig_intro} (a).
At the test time,
when the attacker performs the trigger action (shown in Fig. \ref{fig_intro} (b)), the backdoor will be activated, causing the system to categorize the \siyuan{performed action} 
into a target class \siyuan{specified} by the attacker, regardless of the actual action.}
This method eliminates the dependence on the model architectures or parameters, and removes the need for precise \siyuan{control over joint movements, }
thus addressing the primary challenges of digital attacks.} 


{While backdoor attacks (BA) and their defenses have been well explored in image classification, adapting these attacks to SAR remains a challenge.}
{The core challenge arises from the unique nature of skeleton data, which differs from pixel-based media.}
Unlike images where perturbations can be precisely applied at the pixel level,
{the skeleton data is characterized by a limited number of joints with degrees of freedom \siyuan{that are considerably fewer than the number of pixels in an image.}
This limited dimensionality and the resultant lack of redundancy restrict the attacks to a constrained subspace, narrowing the potential for subtly executing an attack \cite{diao2021basar,tramer2017space}.}
%
{Secondly, 
\siyuan{for BA,}
the manipulations must be imperceptible,
\siyuan{a criterion that lacks a clear definition in the context of skeleton movements.}
\siyuan{Unlike visual data, where the imperceptibility of perturbations often depends on its magnitude, skeleton motions possess unique dynamics that are keenly detected by human sensory systems.}
Sparse attacks on individual joints or frames, \siyuan{even if minimal, may disrupt the continuity of motion, making the attack noticeable.}
In contrast, synchronized manipulations across several joints and frames can \siyuan{preserve the movement's fluidity,}
staying hidden even with larger alterations \cite{wang2019smart}.}
{Lastly, transitioning from digital to the physical realm necessitates trigger conditions that accommodate the inherent imprecision of human movement.
\siyuan{The execution of actions in the real world exhibits variations across individuals in terms of speed, size, and precision. }
}
{To address this, triggers should factor in human variability, incorporating a degree of tolerance.
}

{
Our approach involves simulating realistic skeleton movements that effectively embed trigger actions, avoiding the need for digital modifications with high precision.
To manifest the trigger action, we manipulate the joint angles within the skeleton chain, considering the dynamics and constraints of physical movement. 
The design of the trigger actions' spatial configurations incorporates considerable tolerance, ensuring their effective activation despite variations in human performance.
We introduce an innovative backdoor attack method that remains covert against existing defenses and is adaptable enough for real-world scenarios where exact replication of joint movements is impractical.}
%
%

\qichenn{To summarize, our contributions are four-fold: 1) To the best of our knowledge, this is the first work considering the BA for SAR. Motivated by the unique properties of skeleton data, we introduce the Physical Skeleton Backdoor Attack (PSBA). 2) We validate PSBA's effectiveness in both clean-label and poison-label scenarios, developing a systematic method for embedding triggers into skeleton sequences. Furthermore, we introduce a trigger-enhancing strategy specifically designed for clean-label settings. 
3) We assess PSBA's robustness against various backdoor defenses and adopt metrics to ensure that the triggers remain undetectable.
4) We collect a real-world dataset to simulate physical attacks.}

\section{Related Works}

\subsection{Skeleton Action Recognition}

Recent progress in SAR~\cite{liu2017skeleton,zhang2017view,yang2021skeleton, ke2017new, li2018co,yang2023self} increasingly utilizes deep networks like Recurrent Neural Networks (RNNs), Convolutional Neural Networks (CNNs), and Graph Convolution Networks (GCNs) for their effective feature learning from skeleton sequences. 
Specifically, RNNs~\cite{du2015hierarchical,liu2017skeleton,zhang2017view} are extensively employed for modeling temporal dependencies and capturing motion features, while CNN methods~\cite{du2015skeleton, ke2017new, li2018co} are employed to transform skeleton sequences into uniform maps for spatial-temporal analysis or to apply temporal convolutions to the sequences.
The realization that the human 3D skeleton can be viewed as a natural topological graph has significantly heightened interest in the application of GCNs for skeleton action recognition.
For instance, \cite{yan2018spatial} presents a spatial-temporal GCN to learn both spatial and temporal patterns from skeleton data. 
\cite{chen2021channel} proposes a Channel-wise Topology Refinement Graph Convolution (CTR-GC), aiming to discern distinct topologies in different channels for skeleton action recognition.
More recently, \cite{chi2022infogcn} proposes InfoGCN, a novel framework that merges an information bottleneck strategy with attention-driven graph convolution to learn context-dependent skeleton topologies.



\subsection{Backdoor Attacks}


Based on attacker capabilities, BA~\cite{gu2017badnets,hammoud2021check,wang2021backdoor,nguyen2020input,nguyen2020wanet,li2021hidden,Yu_2023_CVPR} can be categorized into poisoning-based and non-poisoning-based attacks. 
In poisoning-based attacks~\cite{gu2017badnets,chen2017targeted,li2020invisible,liu2020reflection,li2021invisible,yu2024purify}, attackers manipulate the dataset by inserting poisoned data but have no access to the model training.
In contrast, non-poisoning-based attack methods~\cite{dumford2020backdooring,rakin2020tbt,tang2020embarrassingly,guo2020trojannet,doan2021lira} inject backdoors by modifying model parameters.
{Regarding trigger generation, numerous backdoor attack methods employ consistent and fixed triggers. However, some approaches have advanced to develop sample-specific triggers, \textit{e.g.,} \cite{li2021invisible} generate the triggers using autoencoders.}
%

BA have been extensively explored in various domains, including natural language processing~\cite{chen2021badnl}, and even in closely related tasks like point cloud classification~\cite{li2021pointba,xiang2021backdoor}.
{In point cloud classification, where the input data comprises batches of discrete points, triggers are typically introduced by methods such as adding a spherical object.}
%
{Unlike point cloud data, skeleton data utilized in action recognition features precisely defined data points, each corresponding to distinct physical structures. This is exemplified by several keypoints, such as 25 points representing human joints in NTU RGB+D~\cite{shahroudy2016ntu} and PKU-MMD~\cite{liu2017pku} datasets.
Therefore, trigger generation methods for point cloud classification or other tasks may not seamlessly apply to BA against SAR.}


\section{Methodology}

\subsection{Problem Formulation}

In the context of SAR, consider a training authority tasked with learning a classifier \( f_{\boldsymbol{\theta}}: \mathcal{S} \rightarrow \mathcal{Y} \). 
Here, \( \mathcal{S} \subseteq \mathbb{R}^{N \times T \times 3} \) represents the space of skeleton sequences, \( \mathcal{Y} \) denotes the label space, and $\bm{\theta}$ denote the trainable parameters of $f$. {Here, $T$ and $N$ represent the number of frames and body joints, respectively.}
%
%
%
{The learning process of the classifier \( f_{\boldsymbol{\theta}}\) involves training with a dataset $\mathcal{D}_{train}$.}
\( \mathcal{D}_{\text{train}} \) can comprise both clean and poisoned data, denoted as \( \mathcal{D}_{\text{train}} = \mathcal{D}_{\text{clean}} \cup \mathcal{D}_{poison} \). 
\( \mathcal{D}_{\text{clean}} \) consists of genuine, correctly labeled sequences of skeleton data, while \( \mathcal{D}_{poison} \)
{contains the sequences embedded with trigger actions by the attacker.}
%

{The attacker has two primary objectives: The first is to implant a backdoor mechanism in the classifier. This ensures that any skeleton sequence \( \boldsymbol{S} \), once altered with the trigger action \( \mathbf{V} \), is incorrectly classified into a specific target class \( y_t \in \mathcal{Y} \).
The second goal is to maintain the classifier's accuracy on unperturbed skeleton sequences, thereby preventing detection through diminished performance on a validation set.}
%
The overall optimizations formalized by the attacker's objective are given below:
\begin{equation} 
\begin{split}
& ~~~\max_{\boldsymbol{T}_{\mathbf{V}}(\cdot)} \mathbb{E}_{(\boldsymbol{S},y) \sim \mathcal{C}}\left[\mathbb{I}\left(f_{\boldsymbol{\theta}^{*}}(\boldsymbol{T}_{\mathbf{V}}(\boldsymbol{S}))=y_t\right)\right] + ~\mathbb{E}_{(\boldsymbol{S},y ) \sim \mathcal{C}}\left[\mathbb{I}\left(f_{\boldsymbol{\theta}^{*}}(\boldsymbol{S})=y\right)\right],\\[-3pt]
\text{s.t.} ~~ & \boldsymbol{\theta}^{*} = \arg\min_{\boldsymbol{\theta}} \sum_{(\boldsymbol{S}_i, y_i) \in \mathcal{D}_{clean}} \!\!\!\mathcal{L}(f_{\boldsymbol{\theta}}(\boldsymbol{S}_i), y_i)+\!\!\!\!\!\!\sum_{(\boldsymbol{T}_{\mathbf{V}}(\boldsymbol{S}_i), {y}_t) \in \mathcal{D}_{poison}} \!\!\!\!\!\!\!\!\!\mathcal{L}(f_{\boldsymbol{\theta}}(\boldsymbol{T}_{\mathbf{V}}(\boldsymbol{S}_i)), {y}_t),
\end{split}
\end{equation}
where \( \mathcal{C} \) is the clean skeleton data, \( \mathbb{I}(\cdot) \) is the indicator function, and \( \boldsymbol{T}_{\mathbf{V}}(\cdot) \) is the trigger injection function. The loss function \( \mathcal{L} \), the model architecture \(f_{\boldsymbol{\theta}} \), and other hyper-parameters are chosen exclusively by the trainer.

To achieve these objectives, the attacker modifies the benign samples into poisoned ones and constructs a poisoned subset \( \mathcal{D}_{poison}=\{(\boldsymbol{T}_{\mathbf{V}}(\boldsymbol{S}_i), {y}_t)\}_{i=1}^{N_p} \). 
To enhance the stealthiness of the attacks, 
{it is common practice to adopt a low poisoning rate, represented by the ratio \(\frac{|\mathcal{D}_{poison}|}{|\mathcal{D}_{train}|}\), often below a certain threshold like 10\%.}
Furthermore, if the original label \({y}_i\) in  \(\mathcal{D}_{poison}\) are altered to the target class \(y_t\), the strategy is named poison-label attacks. Otherwise, if poisoned samples are selected from \(y_t\), the strategy is considered clean-label attacks.

\qichenn{The assumptions for BA against SAR can be summarized as: 1) The attacker lacks access to the training process, including model architecture and loss function. 2) The attacker has access to part of the training data. 3) For real-world effectiveness, the poisoned data must represent physically plausible movements.}

\subsection{Trigger Implantation via Physical Movements}

Since skeleton data typically comprises over 20 joints, directly estimating the change for all joints to simulate physical movements can be challenging. Therefore, it is more efficient to manually determine the changes for key joints and estimate the adjustments for the remaining joints based on their topological relationships.
Considering the fixed arm lengths, we propose to estimate the joint positions through the manipulation of joint angles within a humanoid robotic system. 
This system mirrors a kinematic chain with multiple degrees of freedom, akin to the human skeletal structure, which is comprised of interconnected links (bones) and joints in a sequential coordinate frame organization. 

The simulation relies on a skeleton sequence \( \bm{S} = \{S_\tau\}_{\tau=1}^{T}\), where each \(S_\tau\) is a skeleton frame at time \(\tau\).
To facialiate the injection of a trigger action \( \mathbf{V} \) into \( \bm{S} \), we consider the dynamics of the joint chain, and adopt quaternion representation for the rotation of all joints from each skeleton frame $S_\tau$. 

\begin{figure}[t]
\centering
\hspace{1mm}
\begin{minipage}[t]{0.5\textwidth}
\centering
\includegraphics[width=0.9\linewidth]{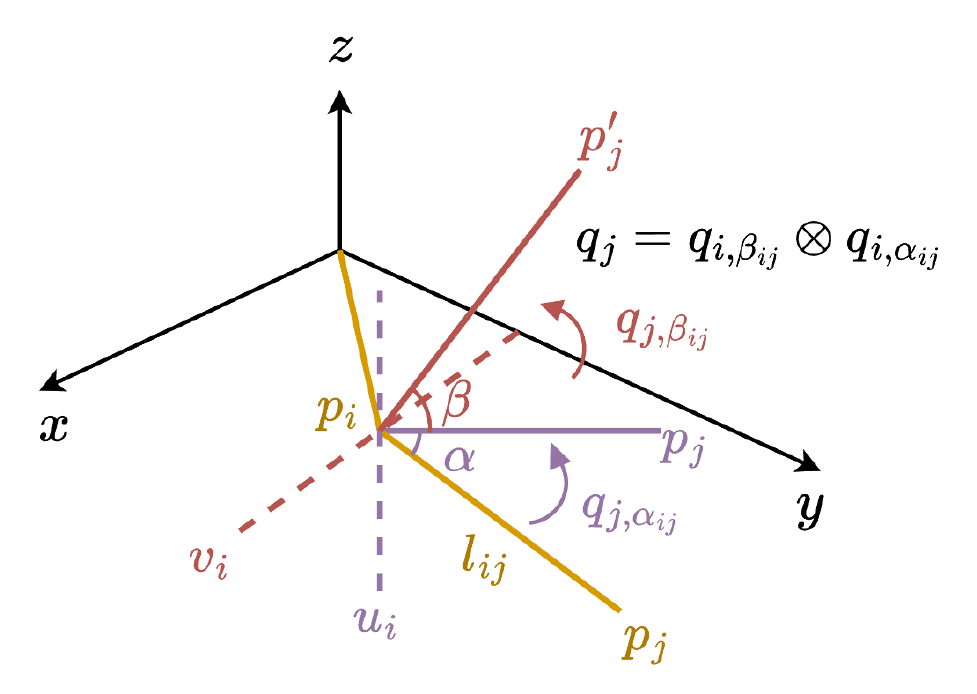}
\centerline{{(a)}}
\end{minipage}
\hspace{1mm}
\begin{minipage}[t]{0.45\textwidth}
\centering
\includegraphics[width=0.83\linewidth]{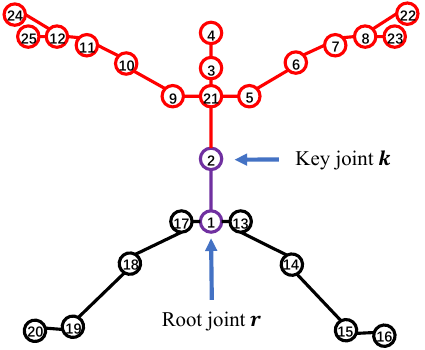}
\centerline{{(b)}}
\end{minipage}

\caption{(a): Schematic diagram of rotation using quaternion. (b): Illustration of joints. \textcolor{violet}{Purple} joints from root joint $r$ to key joint $k$ undergo inverse kinematics, while positions of all colored joints except the root joint $r$ are transformed to the target position in forward kinematics. 
(b) illustrates that when using "bending sideways" as the trigger action, joint 1 serves as the root joint, and joint 2 serves as the key joint.}

\label{Quaternion}
\end{figure}

\setcounter{footnote}{0}

\noindent\textbf{Quaternion representation for rotations.}
Each skeleton sequence frame comprises a set of joints, each defined by its relation to a parent joint and its position in space. As shown in Fig.~\ref{Quaternion} (a), the relationship includes the orthogonal rotational axes vectors \(\bm{u}_i\) and \(\bm{v}_i\) of the parent joint \(i\), the bone length \(l_{ij}\) connecting the joint \(i\) to its child joint \(j\), the rotation angles \(\alpha_{ij}\) and \(\beta_{ij}\) corresponding to rotations around the \(\bm{u}_i\) and \(\bm{v}_i\) axes, respectively, and the position of the joint \(p_j\).
%
To represent rotations as the quaternion \footnote{For the definition of quaternion, please refer to the supplementary material.}, joint $j$ first rotates around $\bm{u}_i$ by $\alpha_{ij}$ angle, and then rotates around $\bm{v}_i$ by $\beta_{ij}$ angle. 
Splitting the rotations into two steps allows the overall rotation process to be expressed as:
 \begin{equation} 
\begin{split}
\!\!\bm{q}_{j,\alpha_{ij}} = \cos(\frac{\alpha_{ij}}{2}) \!+ \sin(\frac{\alpha_{ij}}{2})
&\begin{bmatrix}
\bm{i},\bm{j},\bm{k}
\end{bmatrix}\cdot \bm{u}_i, ~\bm{q}_{j,\beta_{ij}} = \cos(\frac{\beta_{ij}}{2})\! + \sin(\frac{\beta_{ij}}{2})\begin{bmatrix}
\bm{i},\bm{j},\bm{k}
\end{bmatrix}\cdot \bm{v}_i,\\[-4pt]
&~~~~\bm{q}_{j} ={\bm q_{i,\beta_{ij}}}\otimes{\bm q_{i,\alpha_{ij}}},
\end{split}
\end{equation}
where \(\bm{q}_{j,\alpha_{ij}}\), \(\bm{q}_{j,\beta_{ij}}\) is the quaternion for the first and second rotation, respectively. \(\bm{q}_j\) is the quaternion for the overall rotations, and \(\otimes\) is quaternion multiplication. 
%
%
The relative position of joint \(j\) to its parent joint \(i\) is given by:

\begin{equation} 
    \bm{p}_j^{\prime} - \bm{p}_i = \bm{M}(\bm{{q}}_j)\cdot(\bm{p}_j - \bm{p}_i),
\end{equation}
where the matrix $\bm{M}$ is a function of $\bm{{q}}_j$, and is given by:
\vspace{-2mm} 
\begin{equation} 
\bm{M}(\bm{{q}}_j) = \begin{bmatrix}
1 - 2({q}_{yj}^2 + {q}_{zj}^2) & 2({q}_{xj}{q}_{yj} - {q}_{zj}{q}_{wj}) & 2({q}_{xj}{q}_{zj} + {q}_{yj}{q}_{wj}) \\
2({q}_{xj}{q}_{yj} + {q}_{zj}{q}_{wj}) & 1 - 2({q}_{xj}^2 + {q}_{zj}^2) & 2({q}_{yj}{q}_{zj} - {q}_{xj}{q}_{wj}) \\
2({q}_{xj}{q}_{zj} - {q}_{yj}{q}_{wj}) & 2({q}_{yj}{q}_{zj} + {q}_{xj}{q}_{wj}) & 1 - 2({q}_{xj}^2 + {q}_{yj}^2) 
\end{bmatrix},
\label{matrix_M}
\end{equation}
%

\begin{figure}[t]
\centering
\begin{minipage}[t]{0.68\textwidth}
\centering
\includegraphics[width=0.94\linewidth]{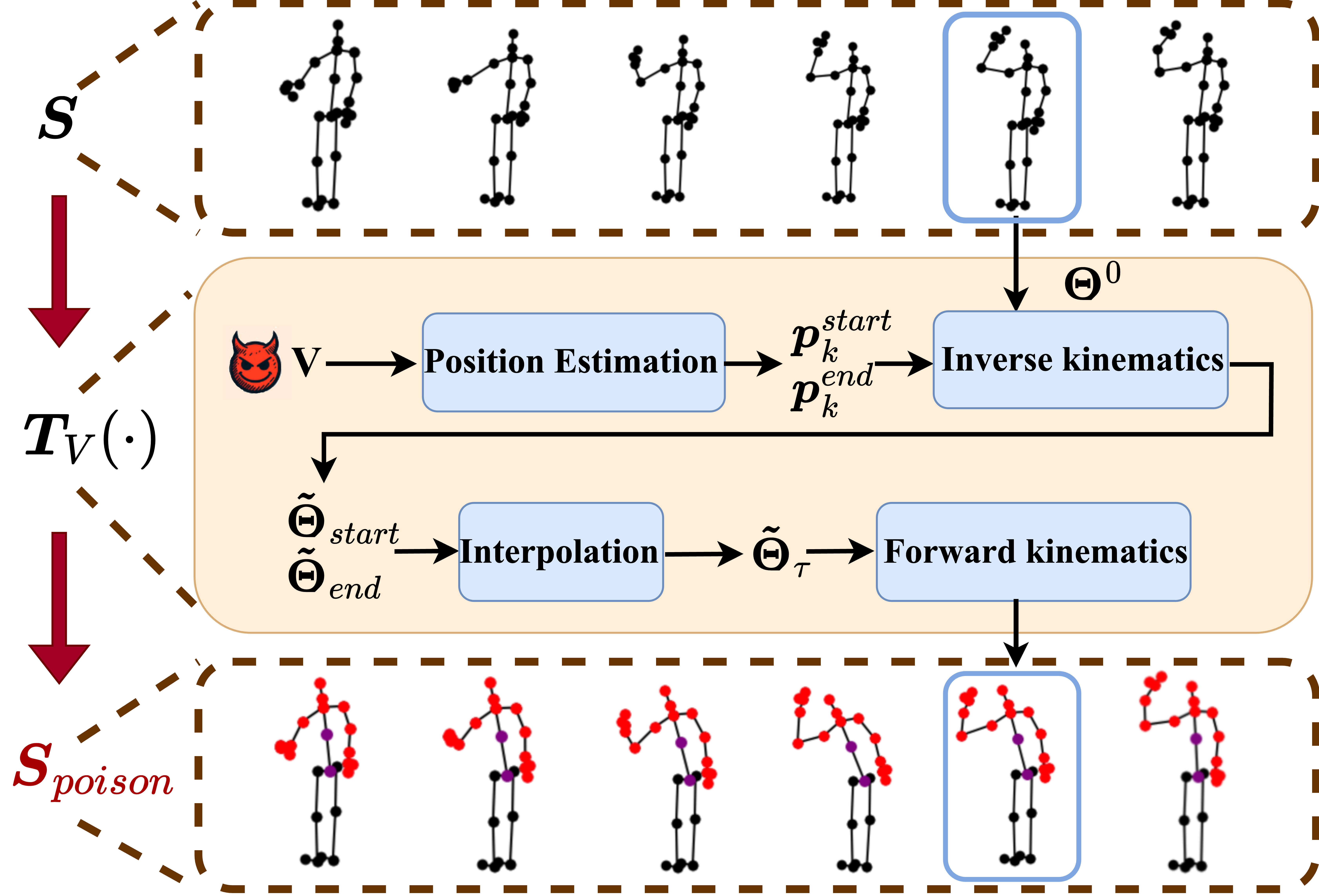}
\centerline{{(a)}}
\end{minipage}
\hspace{-2mm}
\begin{minipage}[t]{0.308\textwidth}
\centering
\includegraphics[width=0.95\linewidth]{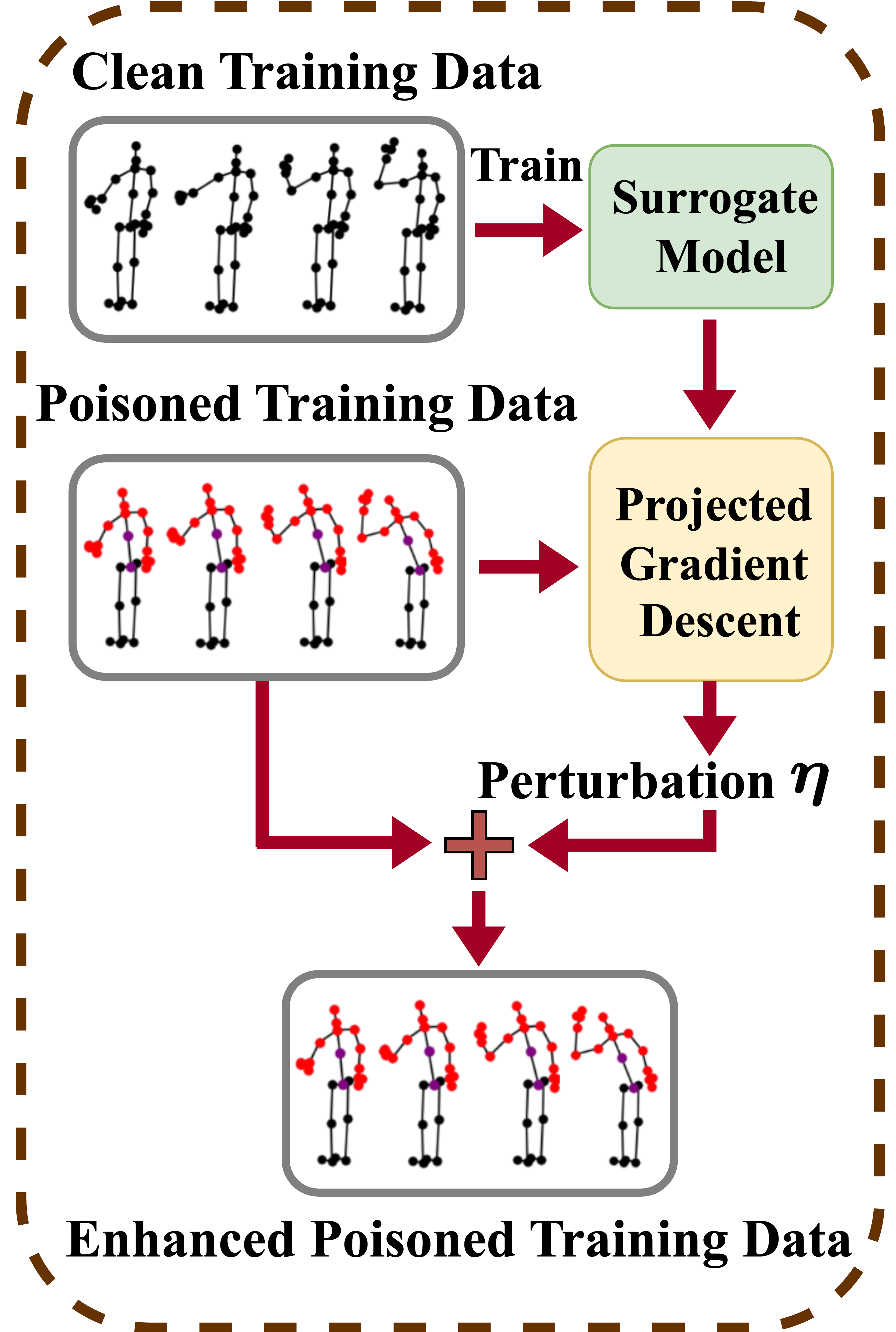}
\centerline{{(b)}}
\end{minipage}

\caption{(a): Trigger implantation diagram. (b): Trigger-enhancing strategy of C-PSBA.}

\label{trigger_injection}
\end{figure}

\noindent\textbf{Poisoned sequences generation via kinematic transformations.}
To generate the poisoned skeleton sequence $\bm{S}_{\text{poison}}$ from the original one $\bm{S}$, we need to calculate the kinematic transformation between these sequences, \siyuan{as shown in Fig.~\ref{trigger_injection} (a)}. 
This transformation associated with any action can be characterized by the variations in joint orientations along the kinematic chain. 

\textbf{Position estimation.}
Given the pre-defined trigger action \( \mathbf{V} \), we first need to define the root joint $r$ that is unchanged, \textit{e.g.,} the base of the spine joint for bending sideways.
Then, we select the key joint $k$ (\textit{e.g.,} the middle of the spine joint for bending sideways), and estimate the corresponding spatial positions for both the start and end state, denoted as $\bm{p}^{\text{start}}_k$ and $\bm{p}^{\text{end}}_k$, respectively.

\textbf{Inverse kinematics from joint $r$ to joint $k$.} Subsequently, we employ inverse kinematics to compute the target orientations for all joints on the chain from the root joint $r$ to the key joint $k$, based on the target position of key joint $k$ estimated in the prior step.
Inverse kinematics is solved using a numerical algorithm such as the Jacobian transpose method, which iteratively adjusts the joint orientations to minimize the difference between the current position and the target position of joint $k$. The iterative optimization can be described by:
\begin{equation} 
\mathbf{\Theta}^{{iter}} = \mathbf{\Theta}^{{iter-1}} + \lambda \cdot \mathbf{K}^T(\mathbf{\Theta}^{{iter-1}})(\bm{p}^{\text{target}}_k - \mathbf{F}(\mathbf{\Theta}^{{iter-1}})),
\label{eq: inverse}
\end{equation}
where $\mathbf{\Theta}^{{iter}}=\left[\alpha_{rm}, \beta_{rm}, \ldots, \alpha_{ek}, \beta_{ek}\right]^T$ is a vector comprising $\alpha_{ij}$ and $\beta_{ij}$ of $\bm{q}_j$ for all joints on the chain from the root joint $r$ to the key joint $k$ at iteration $iter$, $m$ and $e$ are used to represent the child joints of $r$ and the parent joints of $k$, \( \lambda \) is the learning rate, $\mathbf{F}$ is the forward kinematics function that outputs the position of the key joint $k$
as illustrated later, \( \mathbf{K}(\mathbf{\Theta}^{{iter-1}}) \) is the corresponding Jacobian matrix regarding \( \mathbf{F} \) and $\mathbf{\Theta}^{{iter-1}}$, and \( \bm{p}^{\text{target}}_k \) is the desired target position for the key joint $k$. $\mathbf{\Theta}^0$ is initialized by an all-zero vector. 
\siyuan{Therefore, for the spatial positions of both the start and end states, $\bm{p}^{\text{start}}_k$ and $\bm{p}^{\text{end}}_k$, we can utilize the aforementioned inverse kinematics to obtain the desired $\mathbf{\tilde{\Theta}}_{\text{start}}$ and $\mathbf{\tilde{\Theta}}_{\text{end}}$, respectively.
Here, $\mathbf{\tilde{\Theta}}$  represents the target orientations of the final iteration.}

\textbf{Linear interpolation for target orientations.} Let \( \tau_s \) and \( \tau_e \) denote the start and end indices of the action within the sequence, respectively.
\siyuan{For any given frame \( \tau \), utilizing Eq.~\ref{eq: inverse},  we determine the start orientations \( \mathbf{\tilde{\Theta}}_{\tau, \text{start}} \) and the end orientations \( \mathbf{\tilde{\Theta}}_{\tau, \text{end}} \) relevant to frame \( \tau \). 
Subsequently, the desired orientation \( \mathbf{\tilde{\Theta}}_{\tau} \) is calculated by interpolating between the initial and final orientations by:}
%
 \begin{equation} 
\mathbf{\tilde{\Theta}}_\tau = \mathbf{\tilde{\Theta}}_{\tau, \text{start}} + \frac{\tau - \tau_s}{\tau_e - \tau_s} \times (\mathbf{\tilde{\Theta}}_{\tau, \text{end}} - \mathbf{\tilde{\Theta}}_{\tau, \text{start}}).
\end{equation}

\textbf{Forward kinematics from joint $r$ to all affected joints.} As depicted in Fig.~\ref{Quaternion} (b), once we acquire the desired quaternions $\bm{\tilde{q}}$ for the \textcolor{violet}{purple} joints from $r$ to $k$ from $\mathbf{\tilde{\Theta}}_{\tau}$, we can employ forward kinematics to compute the revised positions of all impacted joints, namely, the colored joints excluding the root joint $r$. This process involves computing transformation matrices that define the relationship of each joint's updated position to the original one.

Since rotations accumulate from the parent joint to the child joint, the transformation matrix $\bm{R}_j$ for any joint $j$ (the \textcolor{violet}{purple} one) can be obtained by:
 \begin{equation} 
    \quad \bm{R}_j = \bm{R}_i\cdot \bm{M}(\bm{\tilde{q}}_j), \quad \text{with} ~\bm{R}_r = \bm{I},
\end{equation}
where the matrix $\bm{M}$ is defined in Eq.~\ref{matrix_M}, and joint $i$ is the parent joint of joint $j$.
Note that the parent joint is updated before the child joint.
Thus, the position of any joint $j$ (the \textcolor{violet}{purple} one) can be iteratively updated by:
\begin{equation}
    \bm{p}_j^{\prime} - \bm{p}_i^{\prime} = \bm{R}_j\cdot(\bm{p}_j - \bm{p}_i), \quad \text{with} ~\bm{p}_r^{\prime} = \bm{p}_r,
\end{equation}
where $\bm{p}_j^{\prime}$ is the updated position of joint $j$. 

To simplify, we assume that the distant joints, \textit{i.e.,} the \textcolor{red}{red} ones in Fig.~\ref{Quaternion} (b), have no relative rotation to the key joint $k$. For these \textcolor{red}{red} joint $l$, the transformation matrix $\bm{R}_j$ is the same as $\bm{R}_k$, thus the updated position is:
\begin{equation}
    \bm{p}_l^{\prime} - \bm{p}_k^{\prime} = \bm{R}_k\cdot(\bm{p}_l - \bm{p}_k).
\end{equation}

\subsection{Backdoor Attack Framework}
\label{backdoor_framework}

\noindent \textbf{Physical Trigger Action Design.}
For the trigger actions, we apply the following kinematic transformations and show the selected joints in Table~\ref{affected_joint_action}:
\begin{itemize}

    \item \textbf{Nodding:} This action is characterized by a downward tilt followed by a return to the initial head position. We model this as a two-phase uniform angular sampling for the head joint, first for the nod and then for the return to the upright position.
    \item \textbf{Bending Sideways:} The lateral flexion and subsequent return to the upright position are represented by sampling from a uniform distribution for the spine and hip joint angles, modeling the bend and the return sequence.
    \item \textbf{Crossing Hands at the Front:} This action involves moving the arms from a neutral position to a configuration where hands are crossed in front. The \siyuan{wrist, elbow, and shoulder} joint angles are uniformly sampled to create a smooth transition to this end pose and back. Note that this action involves two separate kinetic transformations \siyuan{(\textit{i.e.,} left and right arms)}.

\end{itemize}

To ensure that the proposed attacks are robust to human variability, the additional design considerations for poisoned samples are summarized as below:
\begin{itemize}

    \item \textbf{Temporal Sampling of Action Duration:} Let the duration range for a trigger action $T$ be a random variable from a uniform distribution $\mathcal{U}(t_{\text{min}}, t_{\text{max}})$ and sample from this distribution to determine the trigger action's timing.
    \item \textbf{Sampling of Hyperparameter for Position Estimation:} For each trigger action, we estimate the $\bm{p}_k^{\text{start}}$ and $\bm{p}_k^{\text{end}}$ of key joint $k$ using a action-related hyperparameter $\Phi$ sampled from a uniform distribution $\mathcal{U}(\phi_{\text{min}}, \phi_{\text{max}})$. For example, $\Phi$ can be the angle to bend for the "bending sideways", and the position of the crossed hands for the "crossing hands at the front".

\end{itemize}

\begin{table}[t]
    \centering
\setlength\tabcolsep{4.5pt}
    \caption{Illustration of the selected joints to affect for each trigger action. }

    \centering
    \scalebox{0.65}{
    \begin{tabular}{c || c c c | c}
    \toprule
    Trigger Action & Root joint $r$ & Key joint $k$ & Joints for inverse kinematics & Remaining affected joints \\
    \midrule
    Nodding & 3 & 4 & 3, 4 & N/A  \\
    \midrule
    Bending sideways & 1 & 2 & 1, 2 & 3 - 12 \& 21 - 25  \\
    \midrule
    \multirow{2}{*}{\shortstack{Crossing hands \\at the front}} & 5 & 8 & 5, 6, 7, 8 & 22, 23  \\
    & 9 & 12 & 9, 10, 11, 12 &  24, 25  \\
    \bottomrule
\end{tabular}}
    \label{affected_joint_action}

\end{table}
\noindent \textbf{Poison-label Backdoor Attacks.}
The poison-label physical skeleton backdoor attack (\textbf{P-PSBA}) is introduced to concretely validate the potency of our designed backdoor triggers. 
{A portion of randomly selected data from the training set undergoes subtle modifications, incorporating trigger action $\mathbf{V}$ into selected skeleton sequences, and the corresponding labels are modified to the target class $y_t$.
When the SAR system is trained with this poisoned dataset, it becomes conditioned to identify these trigger actions as the designated target class.
}

\noindent \textbf{Clean-label Backdoor Attacks.}
{The clean-label physical skeleton backdoor attack (\textbf{C-PSBA}) is designed to bypass label inspection by incorporating triggers without changing the labels.}
Clean-label BA are commonly perceived as the most stealthy strategies, wherein adversaries are restricted to poisoning samples from the designated target class without altering their labels. 

We elucidate that the inherent challenge of clean-label attacks predominantly stems from the adversarial impact of salient features induced by the trigger within poisoned samples. Such salient features are prone to be easily learned, thereby impeding the learning of trigger patterns.
To enhance the efficacy of the trigger and reduce the prominence of original skeleton features, we employ adversarial perturbations, \siyuan{as shown in Fig.~\ref{trigger_injection} (b)}. 
After \siyuan{injecting} the trigger action into selected skeleton sequences from the target class, we introduce subtle adversarial noise into the skeleton data. These perturbations can diminish the original content's impact and encourage the model to focus more on the trigger. 
\siyuan{Notably, }
such perturbations are only added to the poisoned training data, \textit{they are not necessary at run-time}.

Formally, for a given surrogate model \(f_s\) that uses a different model architecture and trained on a clean dataset and an input skeleton sequence \(\bm{S}\) from the target class \(y_{t}\), 
we generate the untargeted adversarial perturbations \(\bm{\eta}\) by maximizing the cross-entropy loss \(\mathcal{L}\) as follows:
\begin{equation} 
\max_{\|\eta\|_2 \leq \epsilon} \mathcal{L}(f_s(\boldsymbol{T}_{\mathbf{V}}(\bm{S}) + \bm{\eta}),y_{t}),
\end{equation}
where \(\epsilon\) is the maximum for \(\bm{\eta}\). 
%
Empirically, we find that such perturbations can transform the original skeleton into hard samples, making it more hard for the model to learn the inherent features associated with the target class. Thus, these perturbations make the learning of the model focus more on the trigger action.

\subsection{Stealthiness}

\qichenn{Stealthiness of poisoned samples within the dataset is crucial to the success of BA, {as it aids in eluding detection.}
To measure the stealthiness from the view of distributions, we employ two statistical metrics: KL Divergence (KLD) and Earth Mover's Distance (EMD).\footnote{Detailed definition can be found in the supplementary material. \label{definition}}}
\qichenn{We calculate the angles between the adjacent bones and statistically distribute on the chain from the joint $r$ to the joint $k$.
By analyzing the distributions of modified dataset and original one, 
we aim to demonstrate that poisoned samples maintain a high degree of stealthiness.}

\section{Experiments}

\subsection{Experimental Setup}

\noindent\textbf{Models.}
{For deep SAR models, we consider a transformer-based and two GCN-based models: Hyperformer~\cite{zhou2022hypergraph}, CTR-GCN~\cite{chen2021channel}, and INFO-GCN~\cite{chi2022infogcn}.

\noindent\textbf{Datasets.}
We select three datasets well-known in SAR.}
\textbf{NTU RGB+D~\cite{shahroudy2016ntu}} 
is a large-scale SAR dataset containing 56,880 skeleton sequences.
Actions are performed by 40 distinct volunteers and are categorized into 60 classes. 
The NTU RGB+D dataset provides two evaluation protocols, namely cross-view (X-view) and cross-subject (X-sub).
\textbf{NTU RGB+D 120~\cite{liu2019ntu}} is an extension to NTU RGB+D, and is the largest SAR dataset, with 114,480 samples over 120 classes. 
\textbf{PKU-MMD~\cite{liu2017pku}} contains almost 20, 000 action instances and 5.4 million frames in 51 action categories, and also utilizes X-view and X-sub evaluation.

\noindent\textbf{Evaluation Metrics.}
\qichenn{To evaluate the effectiveness of our attacks, we utilize two key metrics: Model Accuracy (ACC) and Attack Success Rate (ASR).\textsuperscript{\ref{definition}}}



\noindent\textbf{Attack setting.} For all experiments,
we choose the default class "0" to be the target class (\textit{e.g.,} "drink water" in NTU RGB+D and "bow" in PKU-MMD).

\begin{table}[t]
\caption{ASR (\%) $\uparrow$ and ACC (\%) $\uparrow$ of P-PSBA with different poisoning rates.}

\centering
\scalebox{0.5}{
\setlength{\tabcolsep}{3.0pt}
\begin{tabular}{c|c||*2{c}|*2{c}|*2{c}||*2{c}|*2{c}|*2{c}||*2{c}|*2{c}|*2{c}}
\toprule
\multicolumn{2}{c||}{Dataset $\rightarrow$}&\multicolumn{6}{c||}{NTU RGB+D}&\multicolumn{6}{c||}{NTU RGB+D 120}&\multicolumn{6}{c}{PKU-MMD}\\
\midrule
\multirow{2}{*}{Trigger} & \multirow{2}{*}{\shortstack{Ratio\\(\%)}} &
\multicolumn{2}{c|}{Hyperformer} &
\multicolumn{2}{c|}{CTR-GCN} &
\multicolumn{2}{c||}{INFO-GCN}&
\multicolumn{2}{c|}{Hyperformer} &
\multicolumn{2}{c|}{CTR-GCN} &
\multicolumn{2}{c||}{INFO-GCN} &
\multicolumn{2}{c|}{Hyperformer} &
\multicolumn{2}{c|}{CTR-GCN} &
\multicolumn{2}{c}{INFO-GCN}\\
\cmidrule{3-20}
&& ASR & ACC & ASR & ACC & ASR & ACC& ASR & ACC & ASR & ACC & ASR & ACC& ASR & ACC & ASR & ACC & ASR & ACC \\
\midrule 
  
Clean & - & - & 90.67 & - & 90.08 & - & 90.78 &-&86.62&-&85.08&-&87.36&-&94.85&-&95.23&-&93.58\\
\cmidrule{1-20}
\multirow{6}{*}{Nodding} & 0.1 & 14.72 & 90.35 & 20.34 & 89.69 & 1.71 & 90.55 & 18.90 & 86.57  & 8.33 & 84.99 & 13.65 & 87.42 & 2.51 & 94.49  & 1.33 & 95.03 & 3.67 & 93.27 \\
& 0.2 &  23.98 & 90.38 & 25.48 & 89.94 & 27.39 & 90.55 & 24.61 & 86.45  & 17.57 & 84.96 & 31.22 & 87.12 & 6.94 & 94.57  & 1.83 & 95.01 & 2.33 & 93.35 \\
& 0.5 &  54.75 & 90.46 & 50.55 & 89.96 & 69.49 & 90.49 &58.95 & 86.47  & 60.74 & 84.98 & 64.36 & 87.03 & 13.35 & 94.52  & 2.47 & 95.02 & 4.17 & 92.77 \\
& 1 & 83.30 & 90.41 & 80.28 & 90.33 & 85.50 & 90.86 & 82.84 & 86.39  & 90.26 & 84.96 & 80.12 & 87.13 & 75.98 & 94.61  & 78.52 & 94.98 & 64.09& 92.23 \\
& 2 & 94.26 & 90.48 & 97.58 & 89.94 & 92.75 & 90.64 & 89.39 & 86.37 & 94.18 & 85.01 & 93.17 & 87.27 & 99.50 & 94.39  & 98.67 & 94.96 & 99.33 & 93.12 \\
& 5 & 98.55 & 90.53 & 99.30 & 90.20 & 95.07 & 90.71 & 93.57 & 86.43 & 94.38 & 85.06 & 94.78 & 87.25 & 99.83 & 94.48  & 99.83 & 95.01 & 99.50 & 93.85 \\
\cmidrule{1-20}
\multirow{6}{*}{\shortstack{Bending\\sideways}} & 0.1 & 19.24 & 90.32 & 7.04 & 90.10 & 7.21 & 90.85 & 24.78 & 86.38  & 21.08 & 85.04 & 25.48 & 87.24 & 95.74 & 94.44  & 96.50 & 94.75 & 97.33 & 93.42 \\
& 0.2 & 54.33 & 90.35 & 45.68 & 90.04 & 50.95 & 90.82 & 49.50 & 86.49  & 42.56 & 85.02 & 52.26 & 87.57 & 97.80 & 94.78  & 97.33 & 95.02 & 98.50 & 93.92 \\
& 0.5 & 85.42 & 90.41 & 82.01 & 89.96 & 87.99 & 90.69 & 64.85 & 86.51  & 61.77 & 84.96 & 67.50 & 87.36 & 99.83 & 94.69  & 99.67 & 95.04 & 99.83 & 93.04 \\
& 1 & 88.37 & 90.37 & 85.67 & 89.94 & 92.49 & 90.83 & 87.93 & 86.37  & 91.09 & 84.98 & 88.16 & 87.16 & 99.83 & 94.52  & 99.83 & 94.86 & 99.83 & 92.90 \\
& 2 & 92.88 & 90.54 & 89.91 & 89.88 & 99.20 & 90.88 & 93.14 & 86.43  & 93.78 & 84.99 & 94.88 & 87.31 & 99.83 & 94.63  & 99.83 & 94.90 & 99.83 & 93.38 \\
& 5 & 96.41 & 90.40 & 93.63 & 89.92 & 99.60 & 90.44 & 95.62 &  86.45 & 94.50 & 84.93 & 96.19 & 87.54 & 99.83 & 94.45  & 99.83 & 94.84 & 99.83 & 92.88 \\
\cmidrule{1-20}
\multirow{6}{*}{\shortstack{Crossing \\hands \\at the \\front}} & 0.1 & 29.72 & 90.38  & 22.56 & 89.98 & 24.87 & 90.80 & 5.84 & 86.41  & 1.26 & 85.03 & 0.11 & 87.37 & 26.71 & 94.77  & 24.62 & 94.50 & 20.88 & 93.58 \\
& 0.2 & 57.63 & 90.47  & 39.48 & 89.93 & 44.62 & 90.68 & 11.07 & 86.43  & 2.07 & 84.98 & 0.45 & 87.22 & 28.49 & 94.68  & 22.88 & 95.04 & 28.12 & 93.38 \\
& 0.5 & 82.06 & 90.42  & 76.54 & 89.94 & 84.20 & 90.61 & 88.65 & 86.35  & 86.45 & 84.96 & 94.75 & 87.29 & 68.03 & 94.52  & 81.12 & 94.92 & 57.88 & 93.15 \\
& 1 & 90.85 & 90.39 & 86.10 & 89.78 & 93.68 & 90.73 & 94.39 & 86.29 & 94.87 & 85.01 & 96.99 & 87.18 & 81.19 & 94.56 & 85.50 & 94.88 & 66.00 & 93.73 \\
& 2 & 95.87 & 90.45  & 90.28 & 89.91 & 98.67 & 90.54 & 97.92 & 86.41 & 98.69 & 84.99 & 99.89 & 87.13 & 93.71 & 94.49  & 94.38 & 94.54 & 93.38 & 93.44 \\
& 5 & 98.94 & 90.51 & 94.41 & 89.98 & 99.83 & 90.78 & 99.89 & 86.37  & 99.89 & 85.03 & 99.89 & 87.27 & 94.39 & 94.51  & 94.88 & 95.08 & 94.38 & 93.19 \\
\bottomrule
\end{tabular}
}
\label{table:poison_label_results}

\end{table}

\subsection{Experimental Results}

\noindent\textbf{Poison-Label Attacks (P-PSBA).}
Our experimental analysis unveils the intricate dynamics of BA under different poisoning rates and across multiple model architectures. We conducted evaluations on NTU RGB+D, NTU RGB+D 120, and PKU-MMD datasets, employing poisoning rates of 0.1\%, 0.2\%, 0.5\%, 1\%, 2\%, and 5\%.
The experimental results are shown in Table~\ref{table:poison_label_results}.

\textbf{Influence of Poisoning Rates.}
Our investigation reveals that higher poisoning rates generally elevate the ASR, yet without detrimentally impacting the ACC.
This trend was consistently observed across Hyperformer~\cite{zhou2022hypergraph}, CTR-GCN~\cite{chen2021channel}, and INFO-GCN~\cite{chi2022infogcn}. 
For instance, 
\siyuan{within the NTU RGB+D dataset and using the Hyperformer model under a poison-label scenario, the ASR for the "nodding" trigger action notably escalated from 14.72\% to an impressive 98.55\% as the poisoning rate increased from 0.1\% to 5\%.}
Similarly, other actions such as "bending sideways" and "crossing hands at the front" followed suit, demonstrating the effectiveness of the attacks.
%


\noindent\textbf{Clean-Label Attacks (C-PSBA).}
We do a thorough evaluation under the clean-label setting, applying varying poisoning rates at 30\%, 50\%, 70\%, and 90\% specifically to the data of the target class. 
Experimental result shows that higher poisoning rates can boost the ASR of C-PSBA without negatively affecting the ACC.
\siyuan{In contrast to} attacks in the poison-label scenario,
\siyuan{achieving a high ASR in the clean-label scenario proved challenging due to two primary reasons.}
Firstly, the clean-label approach limits attackers to only poisoning data from the target class, inherently restricting the amount of data that can be manipulated. 
Secondly, in the clean-label scenario, the effectiveness of the trigger pattern can be diminished by the influence of the original features of the target class, \siyuan{which can obstruct the model's learning of the trigger pattern, as discussed in Section \ref{backdoor_framework}.}
%
As shown in Table~\ref{table:exp_clean_backdoor}, although it is possible to attain an ASR between 70\% and 80\% by increasing the poisoning ratio, it proves challenging to achieve an ASR higher than 80\%. 
Thus, attackers must consider the trade-off between stealthiness and effectiveness before launching BA. 
For those aiming for a high ASR, poison-label attacks may be preferable. However, if concerns about label detection exist, clean-label attacks become a viable alternative.

\textbf{Effectiveness of the trigger-enhancing strategy.} 
To assess the efficacy of the trigger-enhancing strategy, we conducted an ablation study on the NTU RGB+D dataset.
\siyuan{As shown in Fig.~\ref{fig_asr_metrics_plot}, our strategy consistently improved the ASR across different poisoning ratios, demonstrating its effectiveness.}

\begin{table}[t]
\caption{ASR (\%) $\uparrow$ and ACC (\%) $\uparrow$  of C-PSBA with different poisoning rates.}

\centering
\scalebox{0.53}{
\setlength{\tabcolsep}{2.9pt}
\begin{tabular}{c|c||*2{c}|*2{c}|*2{c}||*2{c}|*2{c}|*2{c}||*2{c}|*2{c}|*2{c}}
\toprule
\multicolumn{2}{c||}{Dataset $\rightarrow$}&\multicolumn{6}{c||}{NTU RGB+D}&\multicolumn{6}{c||}{NTU RGB+D 120}&\multicolumn{6}{c}{PKU-MMD}\\
\midrule
\multirow{2}{*}{Trigger} & \multirow{2}{*}{\shortstack{Ratio\\(\%)}} &
\multicolumn{2}{c|}{Hyperformer} &
\multicolumn{2}{c|}{CTR-GCN} &
\multicolumn{2}{c||}{INFO-GCN}&
\multicolumn{2}{c|}{Hyperformer} &
\multicolumn{2}{c|}{CTR-GCN} &
\multicolumn{2}{c||}{INFO-GCN} &
\multicolumn{2}{c|}{Hyperformer} &
\multicolumn{2}{c|}{CTR-GCN} &
\multicolumn{2}{c}{INFO-GCN}\\
\cmidrule{3-20}
&& ASR & ACC & ASR & ACC & ASR & ACC& ASR & ACC & ASR & ACC & ASR & ACC& ASR & ACC & ASR & ACC & ASR & ACC \\
\midrule 
  
Clean & - & - & 90.67 & - & 90.08 & - & 90.78 &-&86.62&-&85.08&-&87.36&-&94.85&-&95.23&-&93.58\\
\cmidrule{1-20}
\multirow{4}{*}{Nodding} 
& 30 & 52.20 & 90.34 & 43.31 & 89.90 & 54.28 & 90.53 & 28.42 & 86.34 & 24.30 & 84.96 & 26.74 & 87.26 & 60.20 & 94.63 & 58.06 & 94.98 & 60.98 & 93.56 \\
& 50 & 56.98 & 90.42 & 46.54 & 89.93 & 61.27 & 90.56 & 31.67 & 86.40 & 27.68 & 85.02 & 31.46 & 87.39 & 66.47 & 94.72 & 64.38 & 95.03 & 67.74 & 93.48 \\
& 70& 64.27 & 90.38 & 52.47 & 89.98 & 68.94 & 90.61 & 36.23 & 86.52 & 32.41 & 85.03 & 38.65 & 87.40 & 70.58 & 94.65 & 67.49 & 95.16 & 70.95& 93.36 \\
& 90 & 65.70 & 90.48 & 53.61 & 90.02 & 69.32 & 90.58 & 38.79 & 86.49 & 33.55 & 85.02 & 39.98 & 87.38 & 71.23 & 94.67 & 68.89 & 95.18 & 71.49 & 93.45 \\
\cmidrule{1-20}
\multirow{4}{*}{\shortstack{Bending\\sideways}} 
& 30 & 64.75 & 90.36 & 60.20 & 89.67 & 62.48 & 90.70 & 51.48 & 86.37 & 49.57 & 85.01 & 52.76 & 87.36 & 72.08 & 94.68 & 69.94 & 94.84 & 71.81 & 93.47 \\
& 50& 69.02 & 90.47 & 65.94 & 89.84 & 66.57 & 90.73 & 59.65 & 86.48 & 53.82 & 84.98 & 62.49 & 87.32 & 74.63 & 94.72 & 72.58 & 94.96 & 75.42 & 93.50 \\
& 70 & 75.35 & 90.45 & 71.85 & 90.02 & 74.83 & 90.81 & 65.41 & 86.51 & 60.94 & 85.04 & 68.20 & 87.41 & 77.92 & 94.74 & 74.80 & 95.02 & 78.97 & 93.48 \\
& 90 & 77.14 & 90.39 & 73.04 & 89.79s & 74.95 & 90.79 & 68.93 & 86.59 & 63.02 & 84.98 & 71.44 & 87.42 & 78.36 &94.76 & 75.50 & 95.14 & 79.31 & 93.49 \\
\cmidrule{1-20}
\multirow{4}{*}{\shortstack{Crossing \\hands \\at the \\front}} 
& 30 & 64.34 & 90.46 & 64.21 & 89.84 & 62.48 & 90.63 & 65.72 & 86.44 & 62.30 & 84.86 & 68.65 & 86.92 & 72.83 & 94.69 & 71.18 & 95.08 & 67.50 & 93.37 \\
& 50 & 70.13 & 90.57 & 67.94 & 89.90 & 68.69 & 90.72 & 69.05 & 86.49 & 68.23 & 85.03 & 71.40 & 86.98 & 76.30 & 94.73 & 73.08 & 95.25 & 73.29 & 93.41 \\
& 70& 77.75 & 90.54 & 73.30 & 89.93 & 73.60 & 90.68 & 72.83 & 86.53 & 72.76 & 85.01 & 74.53 & 86.94& 78.51 & 94.79 & 76.43 & 95.21 & 75.69 & 93.45 \\
& 90& 79.06 & 90.52 & 75.92 & 90.04 & 74.51 & 90.84 & 74.91 & 86.52 & 75.30 & 85.06 & 72.61 & 87.01 & 79.88 & 94.78 & 76.94 & 95.21 & 74.84 & 93.42 \\
\bottomrule
\end{tabular}
}
\label{table:exp_clean_backdoor}

\end{table}

\begin{figure}[t] 
  \centering 
    \includegraphics[width=0.7\linewidth]{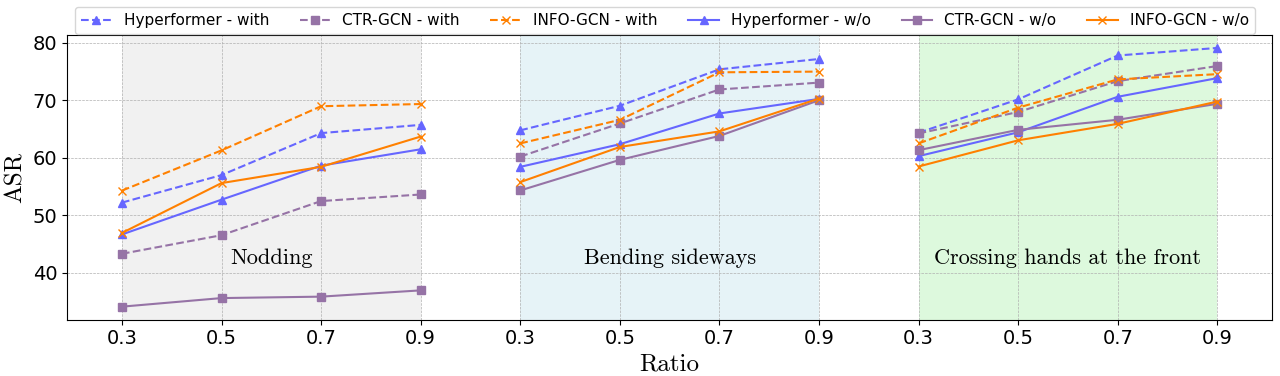}

    \caption{Ablation study on the with/without trigger-enhancing strategy for C-PSBA. 
    }

    \label{fig_asr_metrics_plot}
\end{figure} 

\section{Discussion}

\subsection{Skeleton sequences with trigger from real-world data}

To ascertain the real-world applicability of our proposed BA, we undertook the creation of a physical backdoor attack dataset. \footnote{Further details of the dataset are available in the supplementary.}
We evaluate the practicality of P-PSBA with poisoning rates set at 1\% and 2\%.
As shown in Table~\ref{table:exp_phy_backdoor}, most models trained on the poisoned datasets with a 2\% poisoning rate reliably recognized the embedded triggers with over 85\% ASR, demonstrating the efficacy and reliability of our attacks in the real world.




\begin{table}[t]
\caption{Physical ASR (\%) on models trained with 1\% and 2\% poisoning ratio.}

\centering
\scalebox{0.5}{
\setlength{\tabcolsep}{3.5pt}
\begin{tabular}{c|c||{c}{c}{c}|{c}{c}{c}|{c}{c}{c}}
\toprule
\multicolumn{2}{c||}{Dataset $\rightarrow$}&\multicolumn{3}{c|}{NTU RGB+D}&\multicolumn{3}{c|}{NTU RGB+D 120}&\multicolumn{3}{c}{PKU-MMD}\\
\midrule
{Trigger} & {\shortstack{Ratio (\%)}} &
{Hyperformer} &
{CTR-GCN} &
{INFO-GCN}&
{Hyperformer} &
{CTR-GCN} &
{INFO-GCN} &
{Hyperformer} &
{CTR-GCN} &
{INFO-GCN}\\
\midrule
\multirow{2}{*}{Nodding} 
& 1 &41.87&35.57&39.73&30.68&25.47&29.08&50.14&47.76&49.11\\
& 2 &71.40&66.83&70.62&55.96&49.57&60.71&80.17&70.98&77.79 \\
\cmidrule{1-11}
\multirow{2}{*}{\shortstack{Bending\\sideways}} 
& 1  &81.62&69.83&77.81&69.94&67.66&66.34&83.01&80.19&81.24 \\
& 2 &91.72&88.73&90.02&89.39&86.03&88.57&95.05&90.77&92.13 \\
\cmidrule{1-11}
\multirow{2}{*}{\shortstack{Crossing \\hands}} 
& 1  &71.82&66.91&70.41&61.60&56.55&57.85&75.53&68.35&72.65 \\
& 2  &90.64&81.93&87.72&84.77&82.79&83.59&92.75&89.34&90.48 \\
\bottomrule
\end{tabular}
}
\label{table:exp_phy_backdoor}

\end{table}

\subsection{Resistance to Defenses}

\siyuan{To mitigate the effects of BA in backdoored models, methods range from trigger-synthesis based methods~\cite{wang2019neural,chen2019deepinspect,qiao2019defending}, which synthesize potential triggers and suppress their effects, to solutions like finetuning~\cite{chen2022effective}, pruning~\cite{zheng2022data,liu2018fine}, and input detection~\cite{gao2019strip}.
However, trigger-synthesis based methods are not compatible with our attacks.
Therefore, we select three defenses: CLP~\cite{zheng2022data}, D-BR~\cite{chen2022effective}, and STRIP~\cite{gao2019strip} with their default settings.
Experiments are conducted for P-PSBA using NTU RGB+D as the dataset, INFO-GCN as the SAR model, and "bending" as the trigger action.
For CLP and D-BR, we explore poisoning rates of 0.5\%, 1\%, and 2\% , while STRIP is evaluated at a 2\% poisoning rate.~\footnote{{Additional experiments on defenses are available in the supplementary material.}}
}

\noindent\textbf{Resistance to Prunning based Defenses.}
CLP~\cite{zheng2022data} argues that channels related to BA exhibit higher Lipschitz constants \siyuan{compared to} normal channels. 
By assessing the Lipschitz constant across channels, CLP aims to identify and prune these sensitive channels. 
According to the results presented in Table ~\ref{table:defense_table}, CLP does not effectively counter our P-PSBA approach.

\noindent\textbf{Resistance to Fine-tuning based Defenses.}
D-BR~\cite{chen2022effective} consists of two modules: the Sample-Distinguishment (SD) module and the Backdoor Removal (BR) module. 
The SD module splits the training set into clean, poisoned, and uncertain samples. 
The BR module then alternatively unlearns the distinguished poisoned samples and learns the distinguished clean samples.
\siyuan{We finetune 15 epochs on the NTU RGB+D dataset, and the experimental results are shown in Table ~\ref{table:defense_table}. It can be observed that D-BR is ineffective for our P-PSBA.}


\noindent\textbf{Resistance to Sample Filtering based Defenses.}
STRIP~\cite{gao2019strip} proposes to deliberately inject strong perturbations into each input to effectively identify backdoor inputs.
\siyuan{By analyzing the entropy in the prediction probabilities, it distinguishes between backdoor inputs, characterized by consistently low entropy, and clean inputs, which exhibit high entropy.}
%
\siyuan{As shown in Fig.~\ref{resistance_strip}, the entropy distributions for both clean and poisoned samples in our P-PSBA are similar, suggesting that they can bypass the STRIP defense mechanism.}

\begin{table}[t]
\parbox{.55\linewidth}{
\caption{Resistance of P-PSBA to CLP and D-BR. We set "bending" as the trigger action.}
\centering
\scalebox{0.6}{
\setlength{\tabcolsep}{10pt}
\begin{tabular}{c  c||*3{c}}
\toprule
\multicolumn{2}{c||}{Defense $\rightarrow$} & None & CLP & D-BR \\
\midrule
\multirow{2}{*}{Model $\downarrow$} & Ratio (\%) $\downarrow$ & \multicolumn{3}{c}{ASR} \\
\midrule
\multirow{3}{*}{Hyperformer} & 0.5 & 85.42 & 85.23 & 84.91 \\
& 1 & 88.37 & 88.19 & 88.67 \\
& 2 & 92.88 & 92.81 & 92.50 \\
\midrule
\multirow{3}{*}{CTR-GCN} & 0.5 & 82.01 & 81.94 & 82.99 \\
& 1 & 85.67 & 85.41 & 85.03 \\
& 2 & 89.91 & 89.88 & 90.16 \\
\midrule
\multirow{3}{*}{INFO-GCN} & 0.5 & 87.99 & 87.83 & 85.03 \\
& 1 & 92.49 & 92.46 & 91.87 \\
& 2 & 99.20 & 99.05 & 99.15 \\
\bottomrule
\end{tabular}
}
\label{table:defense_table}
}
\hspace{1mm}
\parbox{.43\linewidth}{
\centering 
\includegraphics[width=0.8\linewidth]{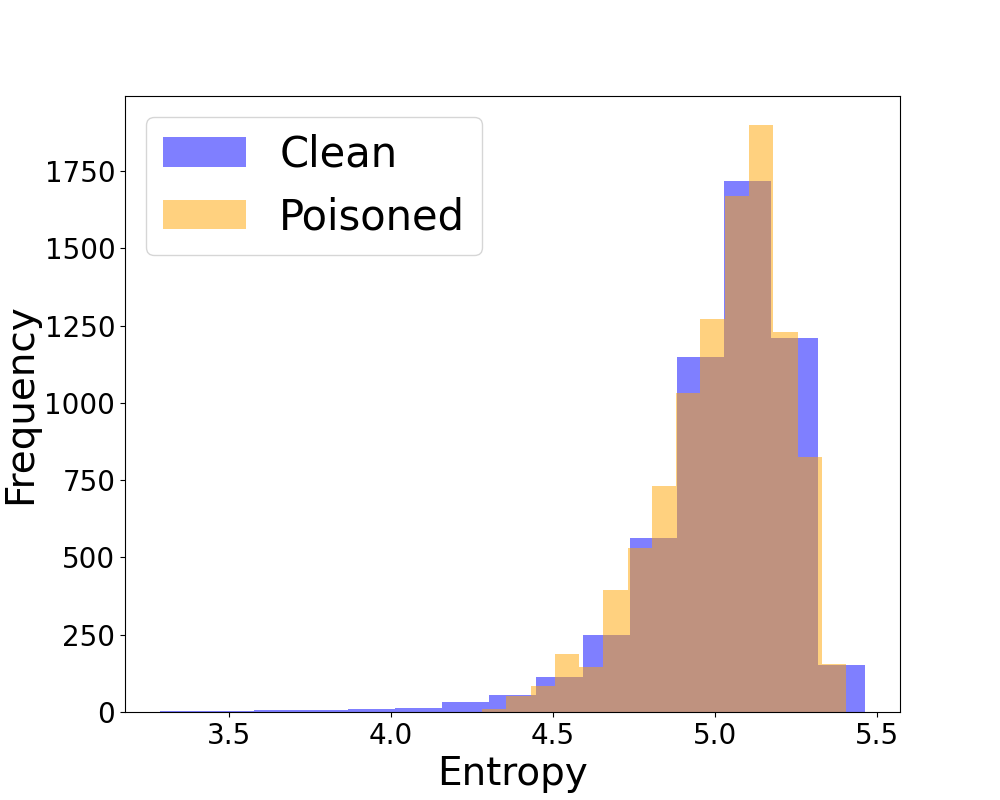}
\captionof{figure}{Resistance of P-PSBA to STRIP. Poisoning ratio is set to 2\%.}
\label{resistance_strip}
}
\end{table}

\begin{table}[t]
\parbox{.55\linewidth}{
\caption{Stealthiness of P-PSBA in terms of KLD $(10^{-7})$ $\downarrow$ and EMD $(10^{-3})$ $\downarrow$.}
\centering
\scalebox{0.6}{
\setlength{\tabcolsep}{5pt}
\begin{tabular}{c||*2{c}|*2{c}|*2{c}}
\toprule
\multirow{2}{*}{\shortstack{Ratio\\ (\%)}} &
\multicolumn{2}{c|}{Nodding} & \multicolumn{2}{c|}{Bending} & \multicolumn{2}{c}{Crossing hands} \\
\cmidrule{2-7}
& KLD & EMD & KLD & EMD & KLD & EMD \\
\midrule 
0.1 & $2.03 $ & 3  & $6.84 $ & 5  & $4.17 $ & 4 \\
0.2 & $4.49 $ & 5  & $16.7 $ & 8  & $7.68 $ & 6 \\
0.5 & $29.9 $ & 11 & $59.4 $ & 16 & $29.2 $ & 11 \\
1   & $108  $ & 21 & $190  $ & 31 & $75.2 $ & 23 \\
2   & $389  $ & 38 & $683  $ & 66 & $256  $ & 48 \\
5   & $2160 $ & 87 & $3930 $ & 139 & $1180 $ & 95 \\
\bottomrule
\end{tabular}
}
\label{table:metric}
}
\hspace{1mm}
\parbox{.43\linewidth}{
\centering 
\includegraphics[width=0.9\linewidth]{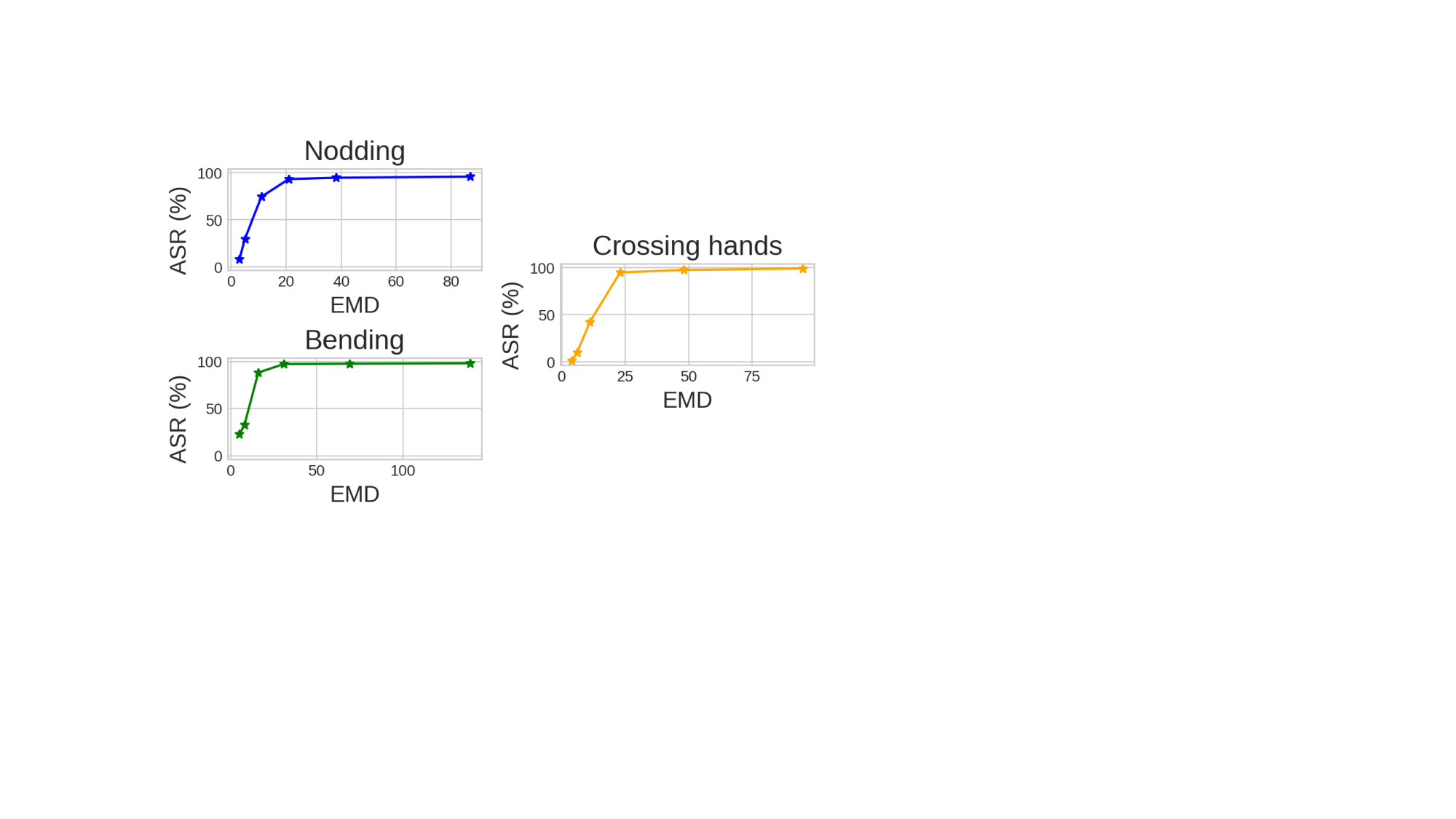}
\captionof{figure}{Plot of EMD $(10^{-3})$ vs. ASR.}
\label{stealthy_asr}
}
\end{table}

\subsection{Stealthiness vs. Attack Performance}

In this section,
\siyuan{the stealthiness of P-PSBA is assessed through the analysis of the KLD and EMD metrics.}
Given the consistency in data processing and \siyuan{integration of backdoor patterns} 
across three datasets, we focus our evaluation exclusively on the NTU RGB+D 120 dataset.
As shown in Table~\ref{table:metric}, despite an increase in poison ratio, both KLD and EMD metrics remained within a relatively low range.
Further, we analyze the relationship between stealthiness and attack performance. As shown in Fig.~\ref{stealthy_asr}, the ASR is highly correlated with EMD metrics. For different triggers, when EMD reaches 20, the ASR can exceed 90\%.

\section{Ethics Statement}

\ysy{This paper proposes a novel backdoor attack against SAR systems. While there is a risk that PSBA could be misused to compromise surveillance systems or cause misclassifications, our objective is to reveal these potential vulnerabilities rather than facilitate attacks. By exposing these vulnerabilities, we emphasize the urgent need for improved defense mechanisms. Our intention is to encourage the development of robust defenses against such attacks, ensuring safer implementations of SAR systems. This work underscores the necessity for the community to recognize backdoor attacks as significant real-world threats and to prioritize their mitigation.}


\section{Conclusion}

In this paper, we introduce a novel backdoor attack specifically tailored for SAR systems, utilizing joint angle manipulations to embed realistic trigger actions.
Comprehensive experiments on both clean label and poison label settings indicate that the proposed attack achieves significant attack success rates and notable resistance against several defenses while maintaining high model accuracy.
Moreover, we design a trigger-enhancing strategy that significantly improves the ASR in clean-label scenarios.
Additionally, our validation using real-world datasets highlights the immediate necessity for advanced defensive strategies in crucial applications.
This work underscores the vulnerabilities in SAR systems and sets the stage for future research into robust defense mechanisms.
\\
\textbf{Acknowledgments}
This research is supported by the National Research Foundation, Singapore and Infocomm Media Development Authority under its Trust Tech Funding Initiative. Any opinions, findings and conclusions or recommendations expressed in this material are those of the authors and do not reflect the views of National Research Foundation, Singapore and Infocomm Media Development Authority.
This research is also supported in part by the NTU-PKU Joint Research Institute and the DSO National Laboratories, Singapore, under the project agreement No. DSOCL22332.


%
%

\end{document}


\title{Towards Physical World Backdoor Attacks against Skeleton Action Recognition (Supplementary Materials)} 

\titlerunning{Physical Backdoor Attacks against Skeleton Action Recognition}

\author{Qichen Zheng\inst{1}\orcidlink{0000-0003-3490-0333} \and
Yi Yu\inst{1}\orcidlink{0000-0003-2730-9553} \and
Siyuan Yang\inst{1}\orcidlink{0000-0003-4681-0431}\thanks{Corresponding author.}  \and
Jun Liu\inst{2}\orcidlink{0000-0002-4365-4165} \and
Kwok-Yan Lam\inst{1}\orcidlink{0000-0001-7479-7970} \and
Alex Kot\inst{1}\orcidlink{0000-0001-6262-8125}}
\authorrunning{Q. Zheng et al.}

\institute{Nanyang Technological University, Singapore \\
\email{\{qichen001,yuyi0010,siyuan005,kwokyan.lam,eackot\}@ntu.edu.sg } \and
Lancaster University, United Kingdom\\
\email{j.liu81@lancaster.ac.uk}
}

\maketitle

\section{Dataset Construction.}
\label{dataset_collect}
\begin{figure}[t] 
  \centering 
    \includegraphics[width=0.6\linewidth]{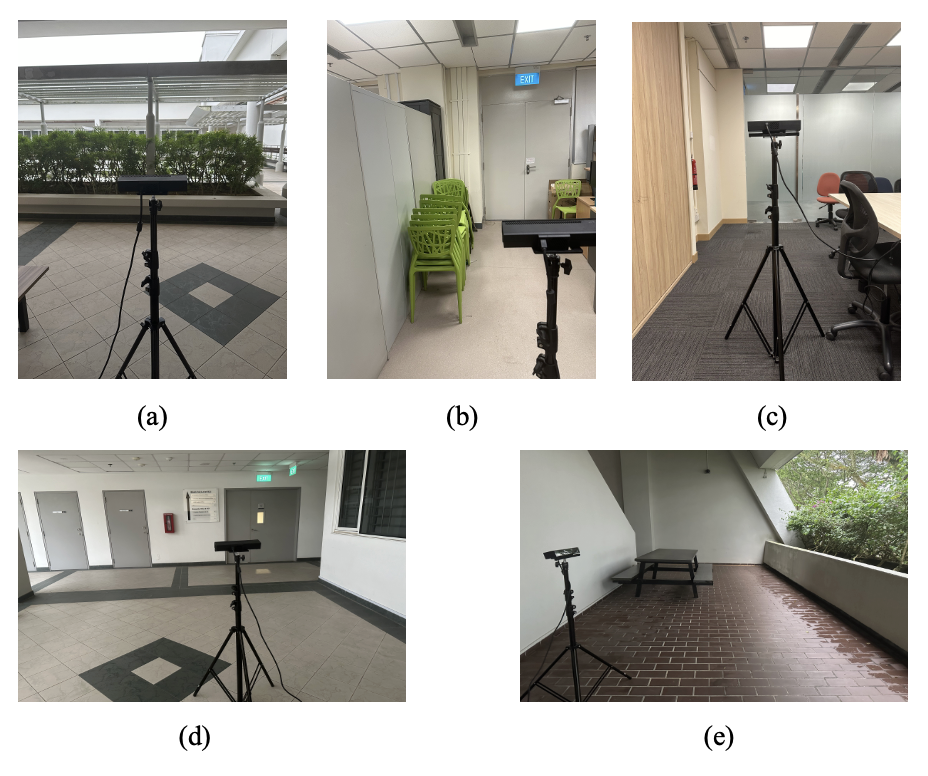}
    \caption{Indoor and outdoor data collection environments.}
    \label{fig_data_env}
\end{figure}

We employed the Kinect V2 camera to capture skeletal data from 3400 instances of action performed in five diverse real-world settings. As shown in Fig.~\ref{fig_data_env}, the real-world scenes include expansive indoor and outdoor environments. 
We recruited 10 volunteers \ysy{to participate in the data collection process.}
Each volunteer was instructed to perform three distinct actions, which have been identified as trigger actions in our experiments: nodding, bending sideways, and crossing hands at the front. 
\ysy{To simulate physical attack scenarios, we analyzed each trigger action performed in conjunction with a separate action.}
We selected 17 common action categories from the NTU RGB+D, NTU RGB+D 120, and PKU-MMD datasets, such as sitting down, jumping up, and kicking, to integrate with the trigger movements.
These actions are enumerated in Table~\ref{tab:trigger_actions}.
\ysy{To incorporate natural human movement variability, each action was executed with varying degrees of motion amplitude—once with a smaller range and once with a larger one.}
Additionally, to ensure a comprehensive capture of the trigger actions, each action was recorded from front and side angles, mitigating potential occlusion issues that could impair the camera's human pose estimation accuracy.
The recorded trigger actions were set aside as test data to assess the model's recognition capabilities. 
\ysy{This assessment is crucial for determining whether our attack methodology can transcend the digital realm and maintain its effectiveness in real-world scenarios.}

\begin{table}[h]
\centering
\caption{List of selected trigger actions}
\scalebox{0.8}{
\setlength{\tabcolsep}{3.8pt}
\begin{tabular}{ccl}
\toprule
Action ID (NTU RGB+D and NTU RGB+D 120)& Action ID (PKU-MMD)  & Action Category \\
\midrule
A6  &A22& Pick up \\
A7  &A41& Throw \\
A8  &A33& Sit down \\
A9  &A34& Stand up \\
A10  &A6& Clapping \\
A11 &A5& Cheer up \\
A14 &A48& Put on jacket \\
A23 &A13& Hand waving \\
A24  &A19& Kicking something \\
A26  &A15& Hopping \\
A27 &A17& Jump up \\
A31 &A25& Point to something \\
A33 &A4& Check time (from watch) \\
A34 &A31& Rub two hands \\
A37 &A50& Wipe face \\
A38 &A32& Salute \\
A40 &A7& Cross hands in front \\
\bottomrule
\end{tabular}
}
\label{tab:trigger_actions}
\end{table}

\section{Implementation Details for Clean Label Attack.}
\label{imp_clean}
We selected the widely used ST-GCN\cite{yan2018spatial} as the surrogate model in the trigger-enhancing strategy. 
After incorporating the trigger action into selected skeleton sequences from the target class, the trigger-enhancing strategy introduces slight, untargeted adversarial perturbations to the skeleton data. 
\ysy{The objective of these perturbations is to diminish the influence of the original content, thereby prompting the model to focus more on the trigger.}

To generate appropriate perturbations without compromising the integrated trigger actions, we constrain the perturbations to the lengths of the skeleton’s bones, following \cite{tanaka2022adversarial}.
\ysy{Additionally, we set the number of iteration steps for the Projected Gradient Descent (PGD) algorithm to five.}

\section{User Study for Trigger Imperceptibility}
\label{us_trigger}
To evaluate the imperceptibility of the trigger action, we conducted a user study
\ysy{in which participants watched a video consisting of 200 skeleton sequence clips derived from the NTU RGB+D test set,}
with some clips containing inserted triggers. 
Participants recorded the index of actions that appeared problematic based on their synthetic likelihood or physical implausibility. 
Indexes were prominently displayed in the top-left corner of each clip for easy reference.

Participants began the study by viewing 10 clean samples without any triggers, serving as a ground truth for comparison. 
Subsequently, they were expected to identify clips that met the evaluation criteria, focusing on their likelihood as triggered sequences.
The result of this study revealed that only 7.1\% of the poisoned samples were identified by the volunteers, underscoring the difficulty in detecting our trigger. 
Additionally, we analyzed the correlation between Earth Mover's Distance (EMD) and the frequency of samples identified in the user study.
As shown in Table~\ref{table:exp_freq_EMD}, there is a positive correlation, indicating that our metrics can effectively measure the stealthiness of the attack to some extent

\begin{table}[t]
\caption{Correlation between EMD and the identified frequency of samples.}
\centering
\setlength{\tabcolsep}{2.9pt}
\begin{tabular}{c||c}
\toprule
Frequency & EMD $(10^{-3})$
\\
\midrule 
 1\% -- 3\% & 3 \\
\cmidrule{1-2}
4\% -- 6\%& 8 \\
\cmidrule{1-2}
7\% -- 9\%& 17 \\
\bottomrule
\end{tabular}
\label{table:exp_freq_EMD}
\end{table}

\section{Metrics and Mathematical Concepts}

\subsection{Metrics}

\subsubsection{KL Divergence (KLD)}
For discrete probability distributions $P$ and $Q$ defined on the sample space $\mathcal{X}$, the KL divergence between $P$ and $Q$ is given by:

\begin{equation}\footnotesize
D_{KL}(P \parallel Q) = \sum_{x \in \mathcal{X}} P(x) \log\left(\frac{P(x)}{Q(x)}\right).
\end{equation}
\subsubsection{Earth Mover's Distance (EMD)}
EMD between two distributions $P$ and $Q$ is the solution to the following optimization problem:
 \begin{equation}\footnotesize 
\text{EMD}(P, Q) = \inf_{\gamma \in \Gamma(P, Q)} \mathbb{E}_{(x, y) \sim \gamma}[\|x - y\|],
\end{equation}
where $\Gamma(P, Q)$ is the set of all possible joint distributions $\gamma(x, y)$. $\text{EMD}(P, Q)$ represents the cost of the optimal transport plan to morph $P$ into $Q$.

\subsubsection{Model Accuracy  (ACC)}
Model Accuracy
{evaluates the classifier's ability to accurately identify clean test sequences. Backdoored models should exhibit high accuracy on clean sequences, aligning with the performance of vanilla-trained SAR models.}

\subsubsection{Attack Success Rate (ASR)}
Attack Success Rate 
{assesses the probability that poisoned sequences embedded with the trigger will be misclassified as the target class \( y_t \) specified by the attacker. It measures the attack's effectiveness.}

\subsection{Mathematical Concepts}
\subsubsection{Quaternion}
Quaternion is employed for rotations to avoid gimbal lock and provide a numerically stable approach for 3D rotations. Given a unit vector \(\bm{u} = [u_x, u_y, u_z]^T\) representing the axis of rotation, and an angle \(\theta\), the corresponding rotation quaternion can be expressed as:
\begin{equation}\footnotesize 
    \bm{q} = q_w + q_x\cdot\bm{i} + q_y\cdot\bm{j} + q_z\cdot\bm{k},
\end{equation}
where \(q_w = \cos\left(\frac{\theta}{2}\right)\), \(q_x = \sin\left(\frac{\theta}{2}\right)u_x\), \(q_y = \sin\left(\frac{\theta}{2}\right)u_y\), and \(q_z = \sin\left(\frac{\theta}{2}\right)u_z\).

\begin{table}[t]
\caption{Resistance of P-PSBA to CLP and D-BR. We set "Nodding" and "Crossing hands at the front" as the trigger actions.}
\centering
\scalebox{0.9}{
\setlength{\tabcolsep}{10pt}
\begin{tabular}{c||  c||  c||*3{c}}
\toprule
&\multicolumn{2}{c||}{Defense $\rightarrow$} & \multirow{1}{*}{None} & \multirow{1}{*}{CLP~\cite{zheng2022data}} & \multirow{1}{*}{D-BR~\cite{chen2022effective}}\\
{Trigger$\downarrow$&Model $\downarrow$ & Ratio (\%) $\downarrow$}&ASR&ASR&ASR\\
\midrule 
\multirow{9}{*}{Nodding}&\multirow{3}{*}{Hyperformer~\cite{zhou2022hypergraph}}
&0.5&54.75&54.37&53.94\\
&&1&83.30&83.21&83.47\\
&&2&94.26&93.98&92.98\\
\cmidrule{2-6}
&\multirow{3}{*}{\shortstack{CTR-GCN~\cite{chen2021channel}}}
&0.5&50.55&50.19&49.75\\
&&1&80.28&80.14&79.64\\
&&2&97.58&97.46&97.23\\
\cmidrule{2-6}
&\multirow{3}{*}{INFO-GCN~\cite{chi2022infogcn}} 
& 0.5 &69.49&69.08&70.02\\
&&1&85.50&85.20&85.13\\
&&2&92.75&92.54&92.28\\
\midrule 
\multirow{9}{*}{\shortstack{Crossing \\hands \\at the \\front}}&\multirow{3}{*}{Hyperformer~\cite{zhou2022hypergraph}}
&0.5&82.06&81.79&81.47\\
&&1&90.85&90.56&90.36\\
&&2&95.87&95.43&95.21\\
\cmidrule{2-6}
&\multirow{3}{*}{\shortstack{CTR-GCN~\cite{chen2021channel}}}
&0.5&76.54&76.31&75.89\\
&&1&86.10&85.73&86.24\\
&&2&90.28&89.92&89.30\\
\cmidrule{2-6}
&\multirow{3}{*}{INFO-GCN~\cite{chi2022infogcn}} 
& 0.5 &84.20&84.01&83.87 \\
&&1&93.68&93.38&93.15\\
&&2&98.67&98.45&98.04\\
\bottomrule
\end{tabular}
}
\label{table:defense_table_supp}

\hspace{1mm}
\end{table}

\section{Resistance to Defenses on more Trigger Actions.}
In the main text, we set "bending sideways" as the trigger action to test our trigger's resilience against backdoor defense methods. In this section, we report the results of two additional trigger actions. We conduct the experiments on the NTU RGB+D dataset in poison-label scenarios. As shown in Table~\ref{table:defense_table_supp}, when selecting "nodding" and "crossing hands at the front" as trigger actions, both CLP and D-BR methods remain ineffective against our P-PSBA approach.


%
%